\documentclass[preprint]{aastex}

\begin{document}
\title{The Evolution of Cas A at Low Radio Frequencies}
\author{J. F. Helmboldt\altaffilmark{1}}
\author{N. E. Kassim\altaffilmark{1}}
\email{joe.helmboldt@nrl.navy.mil}

\altaffiltext{1}{Naval Research Laboratory, Code 7213, 4555 Overlook Avenue SW, Washington, DC 20375-5351}

\received{?}
\accepted{?}

\begin{abstract}
We have used archival 74 MHz VLA data spanning the last 15 years in combination with new data from the Long Wavelength Demonstrator Array (LWDA) and data from the literature covering the last 50 years to explore the evolution of Cas A at low radio frequencies.  We find that the secular decrease of the flux density of Cas A at $\sim80$ MHz is rather stable over five decades of time, decreasing at a rate of 0.7-0.8\% yr$^{-1}$.  This is entirely consistent with previous estimates at frequencies as low as 38 MHz, indicating that the secular decrease is roughly the same at low frequencies, at least between 38 and 80 MHz.  We also find strong evidence for as many as four modes of flux density oscillation about the slower secular decrease with periods of $3.10\pm 0.02$ yr, $5.1\pm 0.3$ yr, $9.0\pm0.2$ yr, and $24\pm2$ yr.  These are also consistent with fluctuations seen previously to occur on scales of a few years.  These results provide compelling motivation for a thorough low frequency monitoring campaign of Cas A to constrain the nature and physical origins of these fluctuations, and to be able to better predict the flux density of Cas A at any given epoch so that it may be used as a reliable low frequency calibrator.

\end{abstract}


\section{Introduction}
Cassiopeia A, or ``Cas A'' is one of the brightest radio sources in the sky, making it one of the most well studied supernova remnants (SNRs).  For some time, it has been known that the radio luminosity of Cas A is fading and that the rate at which it fades varies with frequency \citep[see, e.g.,][]{hog61,sco69,baa77,ree90,osu99,mar02,vin06}.  The canonical expression for the secular decrease of the flux density of Cas A as a function of frequency is that given by \citet{baa77}.  However, recent work has shown that the \citet{baa77} equation may exaggerate the frequency dependence \citep{osu99} and at low ($<100$ MHz) frequencies, the equation significantly overestimates the magnitude of the secular decrease \citep{ree90,mar02,vin06}.  In addition, it has also been demonstrated that this decrease is not smooth, but in fact contains significant variations \citep[see, e.g.,][]{ree90}, usually on time scales of a few years \citep{vin05}.  Placing better and more current constraints on the magnitude and time dependence of both the secular decrease and smaller scale variations of the flux density of Cas A at low frequencies will aid in the establishment of the exact physical causes of these temporal and spectral variations.\par
The implementation of the 74 MHz system for the NRAO Very Large Array \citep[VLA;][]{kas07} in the early 1990s provided a new means to monitor and explore the evolution of Cas A at low frequencies.  New northern hemisphere instruments currently being developed such as the Low Frequency Array (LOFAR) and the Long Wavelength Array \citep[LWA; ][]{kas06} will allow for continued and better monitoring in the future.  We have obtained archival 74 MHz VLA observations as well as new 74 MHz observations with prototype LWA instrumentation of both Cas A and Cygnus A (or, ``Cyg A'') to provide a temporal baseline of more than 15 years with typical spacings of about 1-3 years.  We require Cyg A observations because it is a comparably bright low frequency source with a flux density of 17.35 kJy at 74 MHz \citep{baa77}.  Cyg A is also known to be non-variable because it is a radio galaxy whose flux density at low frequencies is dominated by extended emission spanning several kiloparsecs making any variability on human time-scales negligible.  This makes Cyg A an ideal calibrator/reference source for our efforts.  When combined with other results within the literature, these data have helped us achieve the goal of better constraining the evolution of Cas A at low frequencies up to the current epoch.  In \S 2, we describe the new data and the measurements made with it.  The characteristics of the temporal variations of the flux density of Cas A gleaned using these data are described in \S 3, and the impact this has on our understanding of the SNR and the low frequency flux density scale is discussed in \S 4.

\section{The Data}
All new, unpublished VLA data were selected from the NRAO Data Archive System\footnote{https://archive.nrao.edu/archive/}.  Within the archive, observing runs that used the VLA and the 74 MHz or ``4-band'' system (exact central $\nu=73.8$ MHz) which included observations of both Cas A and Cyg A were searched for.  We identified seven different data sets from six observing programs with a sufficient number of scans of each source to obtain a calibrated total flux density of Cas A using a ``clean component'' model of Cyg A on the \citet{baa77} scale (currently available at http://lwa.nrl.navy.mil/tutorial/) which approximates the 74 MHz intensity of Cyg A with a series of 697 delta functions (or, clean components) in the image plane.  To increase the time and $u\mbox{-}v$ coverage, the two data sets from the same observing program in 1991 (VLA project AK272) were combined into a single data set within AIPS.  We also added to the archival data calibrated visibility data from the VLA Low Frequency Sky Survey \citep[VLSS; VLA project AG735;][kindly provided by A. Cohen]{coh07}, of a field that contains Cas A.  Unlike the other observations, Cas A was not in the center of the field of view for this VLSS field, and the flux density obtained from these data was corrected for the response of the primary beam of the VLA antennas.\par
To obtain an integrated flux density of Cas A from each data set, the data were first calibrated with standard AIPS routines using the aforementioned clean component model of Cyg A.  Specifically, the AIPS routine CALIB was used with the Cyg A model to solve for the complex gain for each VLA antenna and each scan of Cyg A within time intervals of 1-3 minutes, depending on the data set.  The CALIB task also flags time steps at which no good solution can be obtained.  Following this, the gains determined by CALIB were applied to the scans of Cas A using the task CLCAL to interpolate onto the times of the Cas A scans.\par
Following this, the visibility data were tapered with a Gaussian with a full width at half maximum (FWHM) in the $u\mbox{-}v$ plane that corresponds to a Gaussian with a FWHM of 600 arcsec in the image plane (0.30 kilo-wavelengths), roughly three times the angular diameter of Cas A.  This tapering essentially takes Cas A from a relatively complex extended source and smoothes it to a single Gaussian source.  This implies that any phase errors present in the Cyg A-based calibrations will have a negligible effect on the mean tapered amplitude as a function of $u\mbox{-}v$ radius.\par
\begin{figure}
\plotone{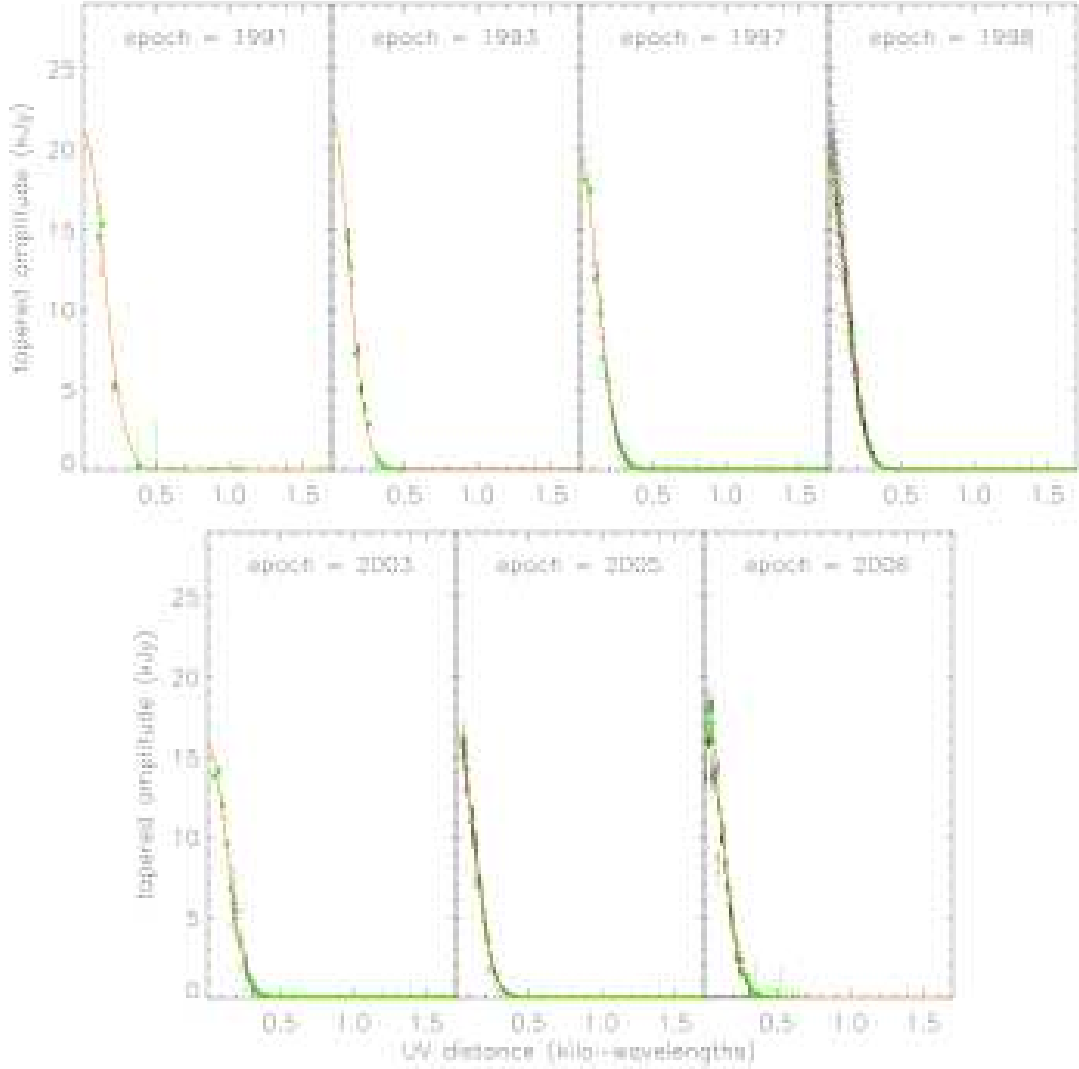}
\caption{For each of the seven epochs for which we have VLA 74 MHz data of Cas A and Cyg A (see \S 2 and Table \ref{obsdat}), the amplitude of the tapered visibilities for Cas A (see \S 2) versus $u\mbox{-}v$ radius.  The data are plotted as black dots with mean values within bins of $u\mbox{-}v$ radius plotted as green dots with the standard deviation within each bin represented by the green error bars.  Gaussian fits to the binned data are plotted in orange.}
\label{radplot}
\end{figure}

Since Cas A is substantially brighter than any other source within the 74 MHz VLA primary beam (the typical total sky flux density at 74 MHz within the VLA primary beam is about 300 Jy; A. Cohen, private communication, 2009) or any sources of RFI, no attempt was made to account for other sources within the field of view or to correct for the effects of RFI.  Instead, the mean amplitude of the tapered data was computed within bins of $u\mbox{-}v$ radius, taking the standard deviation within each bin to be the 1$\sigma$ error which includes the effects of both thermal noise.  The radial amplitude plots, including the binned data, can be seen in Fig.\ \ref{radplot}.  A non-linear least-squares fitting routine was then used to fit a Gaussian function plus a constant to the binned data (see Fig.\ \ref{radplot} for plots of these fits with the data), where the constant allows for the effects of noise, RFI, other sources in the field of view, etc., and the center of the Gassian was fixed at a $u\mbox{-}v$ radius of zero.  The peak amplitude of the fitted Gaussian (i.e., the amplitude at a $u\mbox{-}v$ radius of zero) was then taken to be the total 74 MHz flux density of Cas A at that epoch, and the covariance matrix computed by the fitting routine for the free parameters was used to extract the 1$\sigma$ error in the flux density.\par
\begin{figure}
\plotone{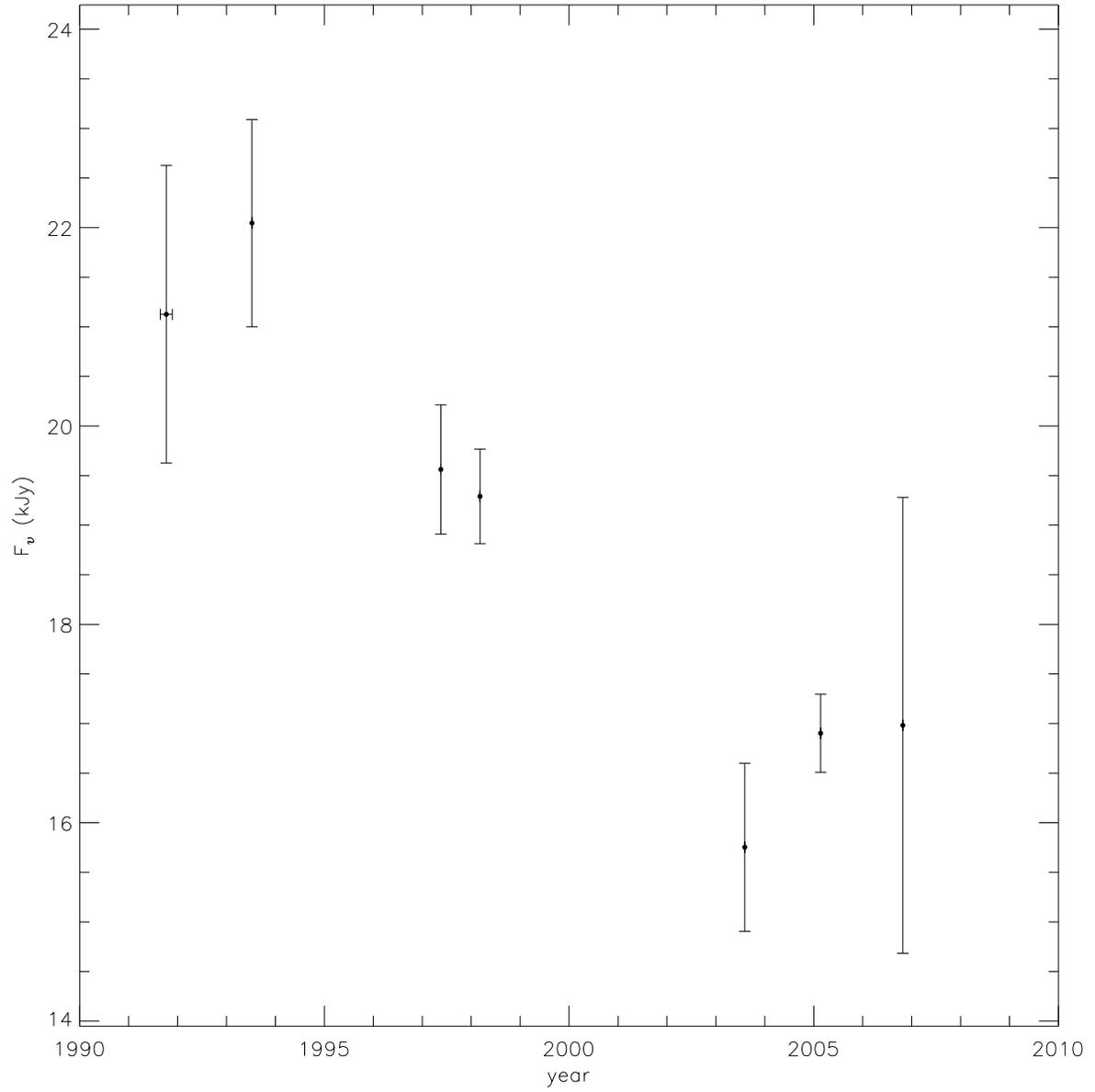}
\caption{The total 74 MHz flux density of Cas A (see \S 2 for a description of the flux and error measurements) as a function of epoch.}
\label{evol}
\end{figure}

We note, however, that this uncertainty may not accurately reflect the true uncertainty in the measurement of the absolute flux density since it is not clear whether or not the variance with each $u\mbox{-}v$ radius bin contains a contribution from the uncertainty in the Cyg A-based calibration.  Given the relative sophistication of the AIPS routines used, simply estimating the uncertainty in the measured Cyg A brightness and propagating this uncertainty through to our final flux density computation is not practical.  However, since each Cas A scan is essentially calibrated separately [i.e., the calibration of a particular Cas A scan is based on the Cyg A scan(s) that is(are) nearest in time], we can obtain a separate estimate of the Cas A flux density uncertainty via jackknife resampling that contains the effects of calibration uncertainty.  We have done this for each data set by re-determining the flux density of Cas A using the same method as described above, each time excluding a different Cas A scan.  The standard deviation among these re-determined flux densities subsequently contains a contribution from the calibration uncertainty since this uncertainty will manifest as an artificial variation in flux density from one scan to the next.  For three of the data sets (AK272, AK332, and AG735), these uncertainties were larger than those obtained from the Gaussian fits displayed in Fig.\ \ref{radplot} and errors computed via jackknife resampling were subsequently adopted as the 1$\sigma$ uncertainties in the total flux densities of Cas A for these data sets.  For all data sets, the flux densities and their final 1$\sigma$ uncertainties are listed in Table \ref{obsdat} and are plotted versus epoch in Fig.\ \ref{evol}.
\par
To supplement these new data, we have also compiled a set of low frequency data from the literature (see below).  Data at 81.5 MHz \citep{hog61,sco69,baa77}, 80 MHz \citep{mar02}, and 38 MHz \citep{ree90,vin06} were obtained and verified to be reliable measurements with no appreciable confusion issues. To add one additional epoch, we have obtained an estimate of the current flux density of Cas A from efforts to measure the beam pattern of antennas being developed for the LWA from September 2008 to February 2009.  Using two LWA prototype antennas on the Long Wavelength Demonstrator Array \citep[LWDA;][]{yor07}\footnote{see the LWA memo series currently located at http://www.ece.vt.edu/swe/lwa/}, \citet{har09} estimated that the flux density ratio of Cas A to Cyg A at 74 MHz over this time period was $1.04\pm 0.05$. We note that all of these additional sources of data either give Cas A flux densities that are on the \citet{baa77} flux density scale, or they directly report Cas A to Cyg A flux density ratios and associated errors.  Therefore, they can be used with our new VLA data to provide a consistent measure of the evolution of the flux density of Cas A.\par

\section{Temporal Variations}
\subsection{Secular Evolution}
It has long been established that the flux density of Cas A has been and is decreasing with time at all radio frequencies \citep[see, e.g.,][]{hog61,sco69,baa77,wal85,ree90}.  This is confirmed by the data plotted in Fig.\ \ref{evol} where the 74 MHz flux density deceases by a factor of about 1.4 between 1991 and 2006.  
Comparing the VLA 74 MHz flux densities to the flux densities used by \citet{baa77} to derive their frequency dependent expression for the secular decrease of Cas A's flux density reveals that the recent VLA-measured flux densities are much higher than what the \citet{baa77} expression predicts.  Specifically, the \citet{baa77} expression estimates a 1.3\% yr$^{-1}$ decrease at 74 MHz.  In contrast, when we combined the Cas A to Cyg A flux density ratios from the data used by \citet{baa77} at 81.5 MHz \citep{hog61,sco69} with our newly presented data, the 74 MHz LWDA data, and the 80 MHz Byurakan Astrophysical Observatory (BAO) data from \citet{mar02}, we found something significantly different.  First, we scaled the Cas A to Cyga A ratios from the VLA and LWDA 74 MHz data and the 80 MHz BAO data to the expected values at 81.5 MHz using the Cas A and Cyg A spectra from \citet{baa77}.  We then performed a non-linear least-squares fit of a secular decrease model to these scaled data and the 81.5 MHz data used by \citet{baa77} and found a smaller decrease of $0.77\pm 0.02$\% yr$^{-1}$ (see the upper panel of Fig.\ \ref{comp}).\par
This is similar to the 0.8\% yr$^{-1}$ decrease at 38 MHz found using data compiled by \citet{ree90}.  In fact, by simply scaling the VLA and LWDA 74 MHz data and the BAO 80 MHz data according to the shapes of the spectra for Cas A and Cyg A given by \citet{baa77}, we have found that both data sets are entirely consistent with the results presented by \citet{ree90}, with the possible exception of the VLA data from 1993.  The same is true for additional 38 MHz available from \citet{vin06}.  This is illustrated in the lower panel of Fig.\ \ref{comp} where we have plotted the scaled 74 and 80 MHz data as well as the \citet{ree90} and \citet{vin06} data with both the \citet{baa77} and \citet{ree90} predictions for the secular decrease of Cas A at 38 MHz.  A new fit to all these data, which is also plotted in Fig.\ \ref{comp}, gives a secular decrease of $0.84 \pm 0.05$\% yr$^{-1}$, nearly identical to that from \citet{ree90}.  It therefore appears that, as noted by \citet{ree90}, the \citet{baa77} expression for the secular decrease of the brightness of Cas A at low frequencies does not hold for recent epochs.  Additionally, the magnitude of this secular evolution appears to be independent of frequency, at least for $38 \leq \nu \leq 80$ MHz, and to not have significantly changed in the last $\sim 50$ years.

\subsection{Smaller Scale Variations}
While the (38 MHz) data plotted in the lower panel of Fig.\ \ref{comp} demonstrate a clear, consistent picture of the secular decrease of the flux density of Cas A, there appears to be some evidence of variations on shorter time scales.  This is indicated by the scatter of the data about the 0.84\% yr$^{-1}$ decrease model which ranges from being within the 1$\sigma$ uncertainties to as much as 2-3$\sigma$.  Specifically, the reduced $\chi^{2}$ ($\chi^{2}$ per degree of freedom) is 1.35.  For uncorrelated, random Gaussian errors, this implies there is a roughly 50\% chance that the simple secular decrease model explains the data by itself.  Assuming these uncertainties are believable, this points to real deviations from the smooth secular decrease of Cas A's brightness which appear to occur on time scales on the order of a few years similar to the variations noted by \citet{vin05}.  To explore this further, we have computed the discrete Fourier transform (DFT) of the 38 MHz data minus the 0.84\% yr$^{-1}$ decrease model.  The mean separation in time between the data points is 1.5 yr and we have therefore only computed the DFT for $| \nu | \leq 0.34 \mbox{ yr}^{-1}$, the approximate Nyquist limit.  The amplitude of the resulting DFT is plotted in the upper panel of Fig.\ \ref{dft} and has 3-6 relatively large peaks.  To explore whether or not the smaller scale fluctuations could be approximated with a sum of regular oscillation modes, we used the properties of the DFT near the six largest peaks as initial guesses for a nonlinear least-squares fit to the data of sums of cosines.  For models with sums of 1, 2, 3, 4, 5, and 6 cosines, the reduced $\chi^{2}$ values (taking into account the two fewer degrees of freedom due to the two parameters of the fitted secular decrease model) were, 1.05, 0.777, 0.762, 0.670, 0.687, and 0.787, respectively.\par
From the $\chi^{2}$-based analysis above, it appears that the smaller scale variations about the secular decrease model are well approximated by a sum of four cosines and that adding more cosines does not improve the fit.  However, we must note that the $\chi^{2}$ quantity is only an adequate measure of the goodness of a fit if the errors used are reliable measurements of the uncertainties in the data.  It is possible that the true uncertainties in several of the data points have been underestimated and that the spectrum seen in Fig.\ \ref{dft} is the result of random variations about the secular decrease of Cas A's flux density that happen to be reproducible with a sum of four cosines.  To check this, we generated $10^{5}$ simulated data sets, each time randomly drawing 37 numbers [i.e., the number of epochs for the combination of the 74 MHz VLA and LWDA data, the 80 MHz BAO data, and the 38 MHz data from \citet{ree90} and from \citet{vin05}] from a normal distribution with a mean of zero and standard deviation equal to the 0.11 rms deviation of the data about the 0.84\% yr$^{-1}$ decrease model.  An examination of the cumulative distribution of these deviations revealed that a normal distribution with a mean of zero and standard deviation of 0.11 approximates the data well.  For each of the simulated data sets, the DFT was computed using the epochs of the real data.  We then used the cumulative distribution of the amplitudes from all $10^{5}$ DFTs at each frequency to compute the 75\%, 90\%, and 95\% confidence levels as functions of frequency.  We have plotted these functions with the DFT of the data in Fig.\ \ref{dft}.  From this plot, it can be seen that in fact, the four largest peaks have amplitudes above the 95\% confidence level which we have highlighted with grey points.  This confirms what the $\chi^{2}$-based analysis implied, that a four-cosine model is a good representation of the real, smaller scale temporal variations in the brightness of Cas A.  The best-fitting four-cosine model parameters and their estimated 1$\sigma$ errors are listed in Table \ref{parms}.\par
We have plotted the (38 MHz) data (minus the 0.84\% yr$^{-1}$ decrease model) in the lower panel of Fig.\ \ref{dft} along with the best fitting four-cosine model.  We have also shaded the area corresponding to the 1$\sigma$ uncertainty in the model which we computed using the covariance matrix of the free parameters and the partial derivatives of the model function with respect to the free parameters at each epoch.  The model appears to represent the data well with the median absolute deviation from the model fit being about $0.8\sigma$.  It can also be seen from Fig.\ \ref{dft} that the model approaches the largest deviant points at epochs of 1961.5, 1967, and 1993.5.  We have also plotted the DFT of the model fit evaluated at the epochs of the data in the upper panel of Fig.\ \ref{dft}.  The main peaks are reproduced well in both location and amplitude while only the locations of some of the smaller peaks are repoduced.  The fact that the amplitudes of the smaller peaks for the DFT of the data are larger is likely due to the influence of noise as well as that of higher frequency oscillations which cannot be fully explored here given the typical sampling rate of our data, but which likely have an effect at lower frequencies due to secondary ``sinc'' peaks caused by the Fourier transform of the 54 yr rectangular envelope of the data (i.e., the data spans epochs from 1955 to 2009).  This points to a distinct possibility that Cas A has (at least) four regular modes of flux density oscillations at low radio frequencies with periods of $3.10\pm 0.02$ yr, $5.1\pm 0.3$ yr, $9.0\pm0.2$ yr, and $24\pm2$ according to the best fitting four-cosine model.

\section{Discussion}
\begin{figure}
\plotone{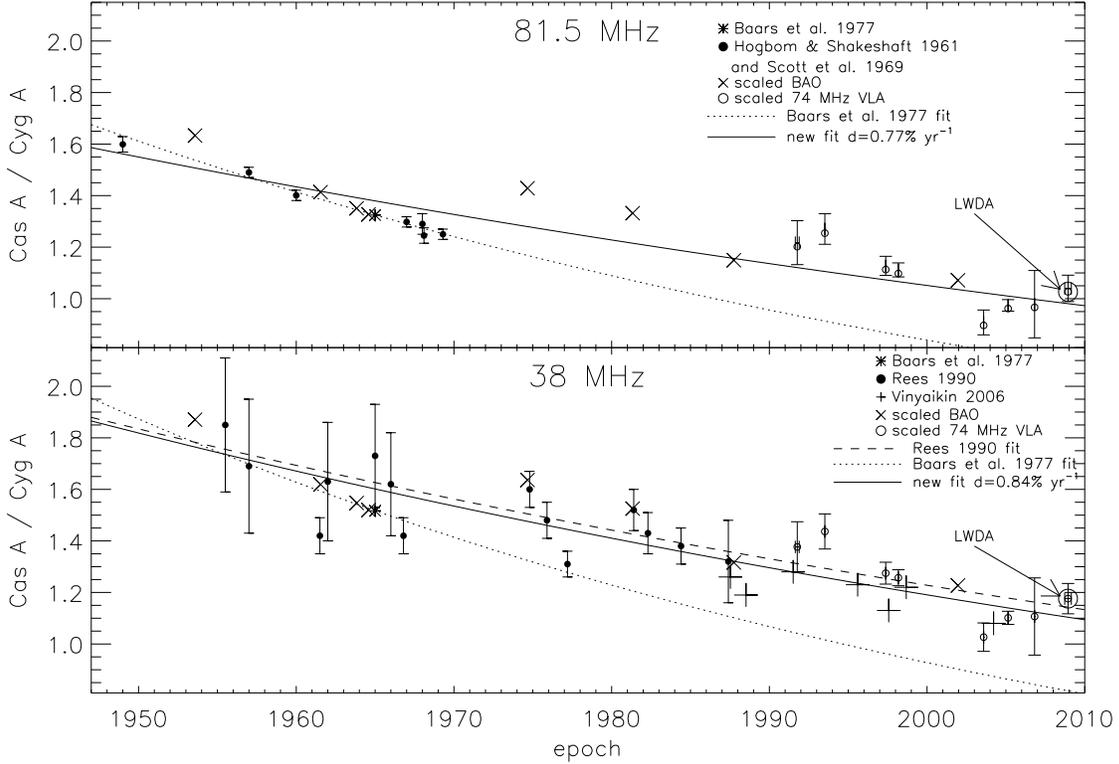}
\caption{Upper: The ratio of the flux density of Cas A to that of Cyg A at 81.5 MHz for our data and other references (see plot legend).  The \citet{baa77} prediction for the secular evolution of Cas A at 81.5 MHz, a 1.3\% yr$^{-1}$ decrease, is plotted as a dotted line; a fit to all the data in the plot is plotted as a solid line, corresponding to a 0.77\% yr$^{-1}$ decrease.  Lower: The ratio of the 38 MHz flux density of Cas A to that of Cyg A for the data compiled by \citet{ree90} with the VLA 74 MHz data scaled to the 38 MHz value assuming the spectral shapes for Cas A and Cyg A given by \citet{baa77}.  Also included is the 38 MHz data of \citet{vin06} and scaled versions of the 80 MHz BAO data of \citep{mar02}.  Again, the \citet{baa77} secular decrease prediction is plotted as a dotted line.  A dashed line is used to represent the 0.8\% yr$^{-1}$ decrease derived by \citet{ree90}; a fit to all the data is plotted as a solid line and corresponds to a decrease of 0.84\% yr$^{-1}$.  For both the BAO and \citet{vin06} data, the data points are roughly the sizes of the uncertainties in the Cas A to Cyg A ratio in both panels.  In both panels, the (scaled) Cas A to Cyga ratio measured with the LWDA antennas by \citet[][see \S 2]{har09} is highlighted with a large open circle.}
\label{comp}
\end{figure}

By combining archival 74 MHz VLA data and the new LWDA data with data from the literature, we have successfully confirmed that the secular decrease of Cas A at low frequencies is in fact not as extreme as the \citet{baa77} results imply.  We have also shown with this data that the low frequency flux density of Cas A oscillates on time scales of $\sim$5-10 years with possibly four or more modes.  These were apparently the cause of the \citet{baa77} misestimation of the magnitude of the secular decrease of Cas A as they did not have data that spanned a long enough time range and unfortunately fit a model to data that included a periodic ``dip'' in the flux density (see the left-most solid points in the upper panel of Fig.\ \ref{comp}).  With data that span 54 years, we have shown that at low radio frequencies, the secular decrease is about 0.8\% yr$^{-1}$, and has remained at this magnitude for five decades.\par
\begin{figure}
\plotone{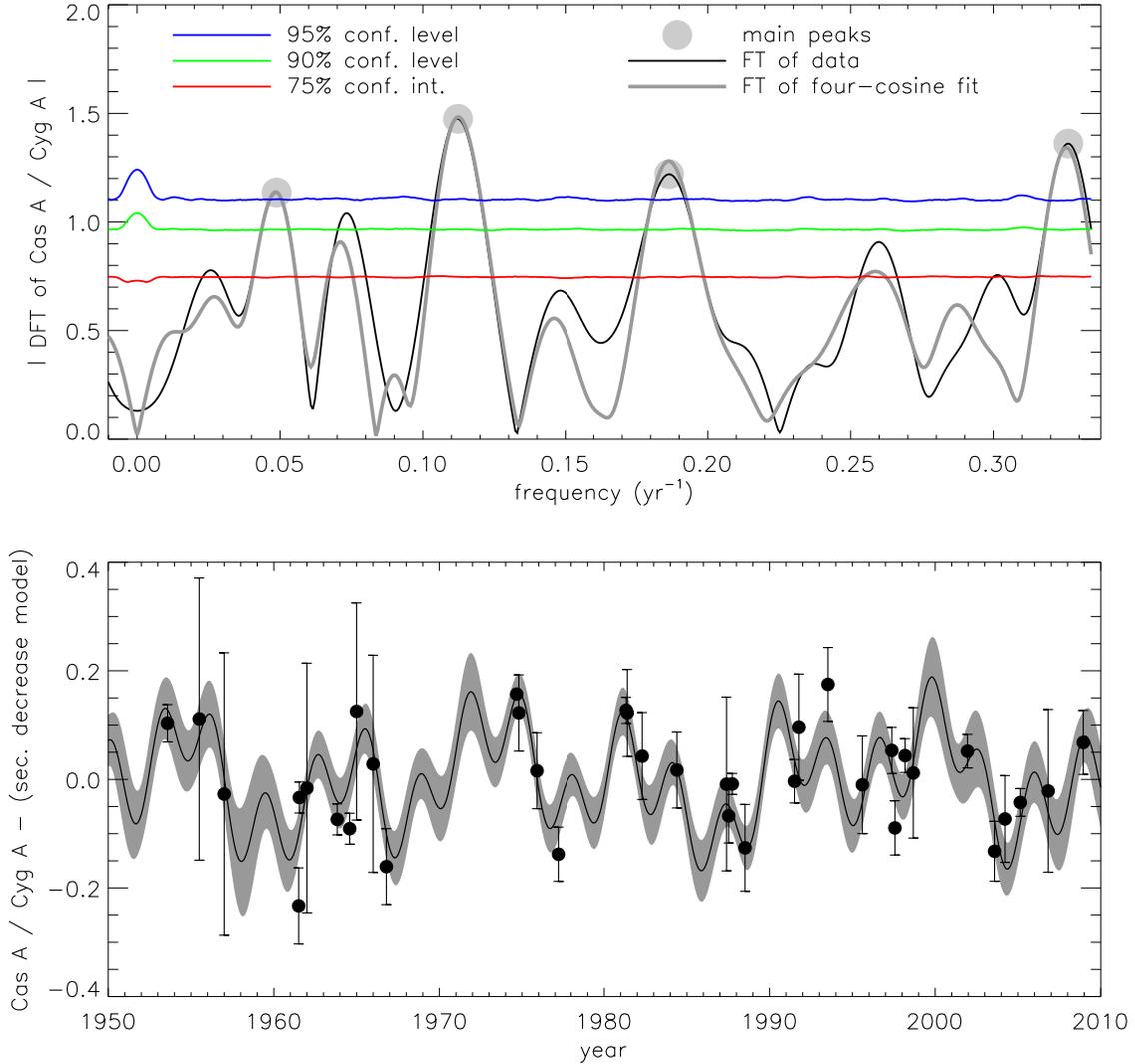}
\caption{Upper: The discrete Fourier transform (DFT) of the data plotted in the lower panel of Fig.\ \ref{comp} minus the best-fitting 0.84\% yr$^{-1}$ decrease model (solid black curve) and the DFT of a four-cosine model fit to the data evaluated at the same epochs as the data (grey curve).  Confidence levels computed at each frequency using Monte Carlo simulations are plotted in red, green, and blue for the 75\%, 90\%, and 95\% levels (see \S 3.2).  The four largest pairs of peaks which are above the 95\% confidence level are highlighted in light grey.  Lower: The data plotted in the lower panel of Fig.\ \ref{comp} minus the 0.84\% yr$^{-1}$ decrease model with a four-cosine model fit plotted as a solid line (see Table \ref{parms} for the model parameters).  The shaded grey area represents the model $\pm 1 \sigma$, where $\sigma$ is the error in the model at a particular epoch based on the covariance matrix of the free parameters and the partial derivatives of the model at that epoch.}
\label{dft}
\end{figure}

The apparent four modes of oscillation we have found are quite interesting.  The DFT of the time series displayed in Fig.\ \ref{dft} shows that the oscillations are in fact very consistent with the sum of four simple periodic functions.  Furthermore, the longer period oscillations seem to be nearly 1:2, 1:3, and 1:8 resonances of the shortest period oscillation.  However, we cannot rule out the possibility that the actual variations are somewhat stochastic in nature and that our four-cosine model model merely provides an adequate approximation at the time resolution of our data.  Still, the parameters of the four-cosine fit may hint at possible mechanisms for the generation of the oscillations themselves, such as the interaction of shock fronts with both ejecta and the external medium; we will leave such issues to be explored by future efforts.\par
The $\sim$5-10 yr oscillations also present a problem for using Cas A as a low frequency calibrator.  While the secular decrease has always made this somewhat problematic, the oscillations worsen the problem.  Over the 51 years spanned by the data, the four-cosine model fit has an rms of about 0.081, corresponding to a variation of 1.24 kJy in the flux density of Cas A from the general secular decrease.  The model fit also suggests that these variations could be as large as $\pm$2-3 kJy.  Our four-cosine model does fit the data significantly better than the secular decrease model alone.  
For instance, when the four-cosine model is combined with the secular decrease model, the data is reproduced to within $\pm$2.7\% on average, with maximum deviations of abut 7\%.  In contrast, the secular decrease model alone does significantly worse, fitting the data within $\pm$5.3\% on average with deviations as large as 15\% or more.  However, the likelihood that these four modes of oscillation will continue unchanged over a long period of time is unknown.  Additionally, over the $\sim$50 year period covered by the data and at the data's time resolution, the four-cosine model may simply be a reasonable approximation of a stochastic process that is difficult to predict.  Therefore, if there are instances where Cas A is needed as a low frequency calibrator, (e.g., 
to calibrate observations of sources which are closer to Cas A on the sky and require phase referencing, or for constraining antenna performance for developing northern hemisphere low frequency arrays such as the LWA and LOFAR), one should be aware that the calibration error will be dominated by the limits in our ability to predict the flux density of Cas A.\par
Overall, it appears that while the secular decrease of Cas A is rather stable at low frequencies over 50 years of observations, we have yet to obtain stringent constraints on the characteristics and nature of the shorter time-scale fluctuations.  While our relatively simple cosine model appears to reproduce the data reasonably well, a more complicated set of oscillation modes, or possibly a more complex model of stochastic variations may be required to predict the flux density of Cas A accurately enough to use it as a low frequency calibrator.  Both this and the study of the physics behind these fluctuations provide ample justification for a more thorough and consistent effort to continually monitor of Cas A at low radio frequencies.

\acknowledgements  The authors would like to thank J. Hartman for kindly providing us with the important results of his work with the prototype LWA antennas and for useful feedback.  We would also like to thank the referee as well as A. Cohen, T. J. Lazio, and W. Lane for helpful comments and suggestions.  This research was performed while the lead author held a National Research Council Research Associateship Award at the Naval Research Laboratory.  Basic research in astronomy at the Naval Research Laboratory is supported by 6.1 base funding.  The National Radio Astronomy Observatory is a facility of the National Science Foundation operated under cooperative agreement by Associated Universities, Inc.

\clearpage
\begin{deluxetable}{cccccc}
\tablecolumns{6}
\tablewidth{0pc}
\tablecaption{Cas A VLA-based Observations and Flux Densities}
\tablehead{
\colhead{} & \colhead{} & \colhead{} & \colhead{$\tau_{\mbox{\scriptsize S}}$} & \colhead{$\Delta \nu$} & \colhead{F$_{\mbox{\scriptsize 74 MHz}}$} \\
\colhead{Obs. Date(s)} & \colhead{VLA Proj. ID} & \colhead{Config.} & \colhead{(min.)} & \colhead{(MHz)} & \colhead{(kJy)} \\
\colhead{(1)} & \colhead{(2)} & \colhead{(3)} &\colhead{(4)} & \colhead{(5)} & \colhead{(6)}}

\startdata
1991-08-25/11-23 & AK272 & A/AB & 131.5 & 1.16 & 21.1$\pm$1.5 \\
1993-07-08 & AK332 & C & 22.8 & 1.16 & 22.0$\pm$1.0 \\
1997-05-17 & AR378 & B & 112.3 & 1.16 & 19.6$\pm$0.7 \\
1998-03-07 & AK461 & A & 21.3 & 1.15 & 19.3$\pm$0.5 \\
2003-08-02 & AD480 & A & 823.5 & 1.15 & 15.8$\pm$0.8 \\
2005-02-19 & AP452 & B & 76.5 & 1.46 & 16.9$\pm$0.4 \\
2006-10-27 & AG735 & C & 5.3 & 1.15 & 17.0$\pm$2.3 \\
\enddata

\tablecomments{Col. (1): Date(s) of the observations.  Col. (2): The project ID within the VLA archive of the data set used.  Col. (3): VLA configuration(s) used.  Col. (4): The total time spent on source (Cas A).  Col. (5): The total observing bandwidth used.  Col. (6): The integrated 73.8 MHz flux density measured using these observations (see \S 2).}
\label{obsdat}
\end{deluxetable}
\clearpage
\begin{deluxetable}{ccc}
\tablecolumns{3}
\tablewidth{0pc}
\tablecaption{Four-cosine Fit Parameters}
\tablehead{
\colhead{} & \colhead{$\nu$} & \colhead{} \\
\colhead{amp.} & \colhead{(yr$^{-1}$)} & \colhead{$\phi$} \\
\colhead{(1)} & \colhead{(2)} & \colhead{(3)}}

\startdata
$0.074\pm0.022$ & $-0.323\pm0.0025$ & $-0.10\pm0.085$ \\
$0.072\pm0.025$ & $-0.111\pm0.0027$ & $0.15\pm0.097$ \\
$0.051\pm0.024$ & $-0.044\pm0.0042$ & $-0.15\pm 0.14$ \\
$-0.019\pm0.026$ & $-0.20\pm0.012$ & $-0.21\pm 0.36$ \\
\enddata

\tablecomments{Each cosine is of the form Cas A / Cyg A = amp cos$[2\pi (\nu t + \phi)]$ where $t$ is time in years; the uncertainties for the listed parameters are the 1$\sigma$ errors from the covariance matrix computed by the fitting routine (see \S 3.2 and Fig.\ \ref{dft}).  Note: these parameters were derived from a fit to the Cas A to Cyg A ratio at 38 MHz.}
\label{parms}
\end{deluxetable}

\clearpage

\clearpage

\clearpage

\clearpage

\end{document}